\documentclass[pra,twocolumn,showpacs,superscriptaddress,preprintnumbers,amsmath,amssymb,floatfix]{revtex4}

\usepackage{graphicx}
\usepackage{dcolumn}
\usepackage{bm}

\begin{document}


\title{Entanglement of group-II-like atoms with fast measurement for quantum information processing}

\author{R. Stock}
\email{restock@physics.utoronto.ca}
\affiliation{Institute for Quantum Information Science, University of Calgary, Alberta T2N 1N4, Canada}
\affiliation{Department of Physics, University of Toronto, Toronto, Ontario M5S 1A7, Canada}
\author{N. S. Babcock}
\affiliation{Institute for Quantum Information Science, University of Calgary, Alberta  T2N 1N4, Canada}
\author{M. G. Raizen}
\affiliation{Center for Nonlinear Dynamics and Department of Physics, University of Texas, Austin, Texas 78712, USA}
\author{B. C. Sanders}
\affiliation{Institute for Quantum Information Science, University of Calgary, Alberta, Canada}

\date{\today}


\begin{abstract}
We construct a scheme for the preparation, pairwise entanglement via exchange interaction, manipulation, and measurement of individual group-II-like neutral atoms (Yb, Sr, etc.). Group-II-like atoms proffer important advantages over alkali metals, including long-lived optical-transition qubits that enable fast manipulation and measurement. Our scheme provides a promising approach for producing weighted graph states, entangled resources for quantum communication, and possible application to fundamental tests of Bell inequalities that close both detection and locality loopholes.
\end{abstract}


\maketitle


\section{Introduction}
Entanglement is a vital resource for most quantum information processing (QIP) tasks, including long-distance quantum communication~\cite{Briegel:1998}, teleportation-based quantum computation~\cite{Gottesman:1999,KLM:2001}, and one-way quantum computation (1WQC)~\cite{Raussendorf:2001}. An under-appreciated but crucial aspect of QIP is the need for speed of single qubit operations, to enable applications including synchronization of quantum communication networks, measurement and feed-forward in 1WQC, and tests of local realism. For example, in 1WQC, the processor speed primarily depends on the time needed for measurement and feed-forward, whereas the entanglement operation may be slow and accomplished simultaneously before commencement of the computation. In atomic systems, single-qubit fluorescence measurements are limited to microseconds due to auxiliary state lifetimes, and in alkali metals single-qubit rotation times are hampered by the gigahertz spectroscopic separations of hyperfine states. In this work, we overcome these obstacles by encoding in long-lived optical clock transitions (e.g., $^1\text{S}_0$ $\leftrightarrow$ $^3\text{P}_0$) of group-II-like neutral atoms, without sacrificing the advantages of other atomic schemes. Group II-like atoms such as Yb and Sr have long been considered for atomic clocks and much recent experimental and theoretical effort has been dedicated to this group of atoms~\cite{Werij:1992,Porsev:1999,Takamoto:2003,Porsev:2004,Takasu:2003,Hong:2005,Barber:2006,Kitagawa:2007}. The recent cooling of Yb into a Bose-Einstein condensate (BEC)~\cite{Takasu:2003} and the ongoing study of interactions~\cite{Kitagawa:2007} make Yb an especially tantalizing candidate for atomic qubits. Our approach for entanglement and measurement of group-II atoms offers promising techniques for the high-speed synchronization needed for quantum communication and computing, and also for the near-term violation of a Bell inequality in a single laboratory, without any assumptions about signaling, sampling, or enhancement~\cite{Clauser:1969,Clauser:1974,Weihs:1998,Rowe:2001}.

Significant experimental progress has been achieved towards entangling atoms in optical
lattices~\cite{nature:insight}, which could lead to the creation of an initial state for 1WQC. Here we take a complementary approach, considering the entanglement of individual pairs of atoms on demand, comparable to other addressable neutral atom architectures~\cite{Dumke:Eckert:Folman}. Rather than creating a generic cluster state, we propose the creation of computation-tailored weighted graph states as a resource for 1WQC and other QIP tasks. Our technique combines efforts to prepare individual atomic qubits from a BEC~\cite{Raizen:all}, coherently manipulate and transport atoms~\cite{Beugnon:2007,Gustavson:2001} using optical tweezers at a ``magic wavelength,'' entangle atoms via an inherently robust exchange interaction~\cite{Hayes:2007,Anderlini:2007}, rotate single qubits via a three-photon optical dipole transition~\cite{Hong:2005}, and perform fast ($\sim$ns) measurements via resonantly enhanced multi-photon ionization (REMPI). A ``loop-hole free'' Bell inequality test imposes stringent requirements on detector separation~\cite{Weihs:1998} and efficiency (see, e.g. the experimental work in~\cite{Rowe:2001,Volz:2006}), and presents an enticing test-bed for fast measurements with applications to QIP. We study the limits of fast measurement for encoding in the optical clock states of Yb and Sr, which can be resolved spectroscopically and measured on a $\sim\!\!10$ns time scale, thereby admitting space-like separation over a few meters (as opposed to large spatial separations considered in~\cite{Volz:2006}). We show that such Bell tests in a single laboratory should be feasible via a detailed theoretical analysis accompanied by comprehensive numerical simulations.


\section{Qubit preparation and transport}
Clock transitions in ions have been considered for effectively encoding qubits for ion trap-quantum computing due to extremely low decoherence rates~\cite{Schmidt:2003,Schmidt:2005}. Similarly, in the case of neutral atoms, optical clock transitions in alkaline-earth and group-II-like atoms are appealing candidates for encoding qubits. Single atoms have been experimentally isolated~\cite{Raizen:all} and transported in optical dipole traps~\cite{Beugnon:2007,Gustavson:2001}. By trapping at a ``magic wavelength''~\cite{Takamoto:2003,Porsev:2004}, the light shift potential is made effectively state-independent, ensuring phase stability of the qubits for several seconds. For example, for the clock states of Sr, the light shift dependencies on the trap laser frequency $\nu$ differ by $d\Delta/d\nu =  2.3\times10^{-10}$~\cite{Takamoto:2003}. Therefore light shift fluctuations can be kept to less than 0.1 Hz by using a trap laser with linewidth of 100~MHz. Furthermore, the magic wavelength at 813.5~nm (easily accessible using commercial lasers) is far detuned from the excited states so that photon scattering rates are on the order of 10~s for trap light intensities of 10~kW/cm$^2$~\cite{Takamoto:2003}. This ensures a coherence time of 10~s or more for trapping and transporting atoms.


\section{Entangling Operation}
We devise a universal entangling operation for bosons, analogous to the recently proposed fermionic spin-exchange gate~\cite{Hayes:2007}. This gate is based on the exchange interaction recently demonstrated for bosonic Rb atoms in a double-well optical lattice~\cite{Anderlini:2007}. Because of inherent symmetrization requirements, gates based on this exchange interaction offer a natural resistance to errors and greater flexibility for encoding atoms, thereby enabling an entangling operation even for atoms with interaction strengths that are state-independent (e.g., Rb~\cite{Anderlini:2007}) or partially unknown, as is the case for most group-II-like atoms (e.g., Yb~\cite{Kitagawa:2007}).

The entangling operation is achieved by temporarily bringing together a pair of atomic qubits via mobile optical tweezers. Unlike state-dependent optical traps wherein atoms are trivially separated into opposite wells after interaction, we have state-independent traps in which the dynamics of the system generally determine the likelihood of a successful separation. However, under adiabatic conditions the atoms definitely end up in opposite wells. We assume a strong confinement to one dimension (1D) by higher order Hermite Gaussian beams according to~\cite{Raizen:all}. All the essential physics is captured in the 1D model we employ here, although performance could conceivably be enhanced by exploiting multi-dimensional effects such as trap-induced resonances~\cite{Stock:2003}.

\begin{figure}[h]
  \centering
  \includegraphics[width=85mm]{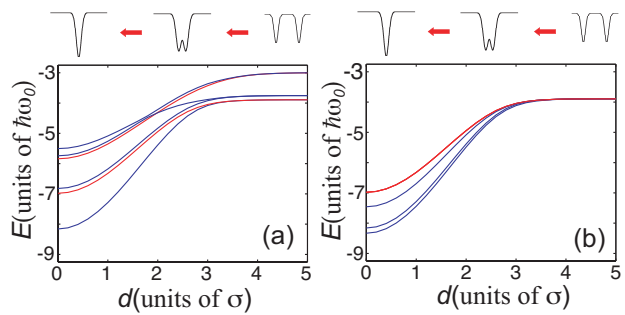}
  \caption{(Colour online) Adiabatic energy levels as a function of well separation. Energies are measured in units of $\hbar\omega_0$, where $\omega_0$ is the harmonic oscillation frequency of one atom in a single well. (a) Lowest six energy levels for $a_{i\! j}=0.1 \sigma$. Energy levels correspond to symmetric or antisymmetric external eigenstates. The antisymmetric curves (red) are the lower of the two curves at $E \approx -7~\hbar\omega_0$ and the lowest of the three curves at $E \approx -5.8~\hbar\omega_0$ for $d=0$. (b) Lowest two levels of (a) for different scattering lengths. The lowest three energy curves (from bottom to top) correspond to $a_{i\! j}=0$, $a_{i\! j}=0.1\sigma$, and $a_{i\! j}=\sigma$, and asymptote to the antisymmetric (topmost) curve for infinite $a_{ij}$. The antisymmetric eigenstates are not affected by the interaction and hence the topmost (red) curve does not shift for different $a_{i\! j}$.
}
\label{fig1}
\end{figure}

The Hamiltonian for two trapped atoms $a$ and $b$ with internal structure ($\left| i
\right\rangle_a,\left| j\right\rangle_b$ $\in\{
|0\rangle,\,|1\rangle\}$) is given by
\begin{equation}
H =\!\sum_{i,j=0,1}\!\!\left[ H_a
+H_b+2a_{ij}\hbar\omega_\bot\delta(x_a\!-\!x_b)
\right]\otimes|ij\rangle\!\langle ij|
\end{equation}
for $H_{a,b}\equiv
p_{a,b}^2/2m+V(x_{a,b}-d/2)+V(x_{a,b}+d/2)$, with
$x_{a,b}$ and $p_{a,b}$ the position and momentum of
atom~$a$ or~$b$. The tweezer potential
$V(x)=-V_0\exp(-x^2/2\sigma^2)$ describes a Gaussian trap of depth
$V_0$ and variance $\sigma^2$. The two wells are separated by a distance $d$, $\omega_\bot$ is the harmonic oscillation frequency of the transverse confinement~\cite{Calarco:2000}, and $a_{ij}$ is the state-dependent scattering length for the two-qubit states $\left|ij\right\rangle \equiv \left| i \right\rangle_a\!\otimes\!\left| j\right\rangle_b$. We numerically solve the Hamiltonian dynamics of individual qubit states using a split-operator method. Two-atom energy spectra are plotted as a function of well separation~(Fig.~\ref{fig1}) for different interaction strengths.

Due to symmetrization requirements, not all combinations of vibrational and qubit states are allowed. For example, a pair of composite bosons cannot share the ground state if the qubits are in the antisymmetric state  $\left|\Psi^-\right\rangle$, defining $\left|\Psi^{\pm}\right\rangle\equiv(\left|01\right\rangle\pm\left|10\right\rangle)\sqrt{2}$. As in the fermionic case~\cite{Hayes:2007}, it is possible to exploit these symmetrization requirements in order to produce a two-qubit entangling operation for bosonic atoms (see \cite{Babcock:2008} for details). Consider a pair of identical bosons, one localized in the left trap ($|\psi^L\rangle$) and carrying a qubit in the state $\left|\varphi^{\alpha}\right\rangle=\alpha\left|0\right\rangle+ \beta\left|1\right\rangle$, the other in the right trap ($|\psi^R\rangle$) and carrying a qubit in the state $\left|\varphi^{\mu}\right\rangle=\mu\left|0\right\rangle+ \nu\left|1\right\rangle$. The initial symmetrized wavefunction (as a tensor product of vibrational and qubit states) is then
$\left|\psi_\text{i}\right\rangle =(|\psi^L\psi^R\rangle\otimes|\varphi^{\alpha}\varphi^{\mu}\rangle
+ |\psi^R\psi^L\rangle\otimes|\varphi^{\mu}\varphi^{\alpha}\rangle)/\sqrt{2}$.

As the wells are brought together and separated adiabatically, the energies evolve as shown in Fig.~\ref{fig1}, and each two-qubit state $|00\rangle$, $|11\rangle$, and $|\Psi^{\pm}\rangle$ acquires a phase  $\phi_{00}$, $\phi_{11}$, and $\phi_{\pm}$, depending on its respective energy curve. Adiabaticity can be satisfied even for negative scattering lengths, since transitions between vibrational states of different symmetry or parity are suppressed. For constant tweezer speed $v$, the adiabaticity criterion is $v \ll {\sigma\hbar\omega_{ab}^2}/{V_0}$. Here, $\hbar\omega_{ab}$ is the energy difference between any coupled states. Time-dependent numerical simulations confirm the validity of the adiabatic approximation over a wide range of values of $V_0$ and $a_{ij}$~\cite{Babcock:2008}. The final state after an adiabatic change of separation is
\begin{align}
\label{bellbasisphase}
    & |\psi_\text{f}\rangle
        =|\psi^-\rangle\otimes(\textstyle\frac{\alpha\nu-\beta\mu}{\sqrt{2}}e^{-i\phi_-}\!\left|\Psi^-\right\rangle) \\
          &\!\!+|\psi^+\rangle\!\otimes\!(\alpha\mu e^{-i\phi_{00}}\!\left|00\right\rangle\!+\beta\nu e^{-i\phi_{11}}\!\left|11\right\rangle\!+\!\textstyle\frac{\alpha\nu+\beta\mu}{\sqrt{2}}e^{-i\phi_+}\!\left|\Psi^+\right\rangle), \nonumber
\end{align}
using $|\psi^\pm\rangle\equiv(|\psi^L\psi^R\rangle\pm|\psi^R\psi^L\rangle)/\sqrt{2}$.

Evidently this process corresponds to a tensor product of the identity acting on the vibrational state and a unitary $U$ acting on the qubit state. Thus, the internal qubit evolution simplifies to
\begin{align}
& \!\!\! U = e^{-i\phi_{00}}|00\rangle\!\langle 00| + \frac{e^{-i\phi_+}\!+\!e^{-i\phi_-}}{2}(|01\rangle\!\langle 01|\!+\!|10\rangle\!\langle 10|) \nonumber \\
& \!\!\! + \frac{e^{-i\phi_+}\!-\!e^{-i\phi_-}}{2}(|01\rangle\!\langle 10|\!+\!|10\rangle\!\langle 01|) + e^{-i\phi_{11}}|11\rangle\!\langle 11|.
\end{align}
As in~\cite{Loss:1998}, a controlled-phase gate can be obtained even if $\phi_+\neq\phi_-$ by sandwiching a single-qubit phase gate between a pair of $U$ operations. That is, $G \equiv U\left[S(\pi)\otimes S(0)\right]U$ for $S(\theta) = \exp(i\theta |1\rangle\!\langle1|)$. Thus defined, $G$ is locally equivalent to $\exp(-i\gamma |11\rangle\!\langle11|)$ if
\begin{equation}
\phi_{00} + \phi_{11} - \phi_+ - \phi_- = (2n\!\pm\!\textstyle\frac{1}{2})\gamma, \quad\forall\;\, n \in \mathbb{Z}.
\end{equation}

As shown in Eq.~(\ref{bellbasisphase}), the phases critical to this entangling operation are acquired in a non-separable basis. This leads to the inherent robustness observed in initial experiments~\cite{Anderlini:2007}. In standard schemes, the important non-separable phase is usually acquired due to the internal state dependence of the interaction strengths $a_{ij}$. In the case of this exchange symmetry-based gate, however, there always is an energy gap between symmetric and antisymmetric curves. The singlet state $\left|\Psi^-\right\rangle$ therefore acquires a phase different from the triplet states even if the interaction strengths are state-independent (except as $a_{ij}\rightarrow\pm\infty$). This substantial phase difference enables the exchange gate to operate faster than standard collisional gates that rely on the difference in $a_{ij}$. Furthermore, this gate works over a large range of scattering lengths [see Fig~\ref{fig1}(b)], which is especially important when designing experiments for atomic species with any currently unknown scattering lengths (e.g., Yb or Sr). Current studies of Yb interactions~\cite{Kitagawa:2007} already promise a wide applicability of this entanglement gate for different isotopes. (For $^{168}$Yb, $a_{00}\approx13$~nm and for $^{174}$Yb, $a_{00}\approx5.6$~nm. $a_{01}$ and $a_{11}$, are not yet known.)


\section{Single qubit rotation and measurement}
\begin{figure}[h]
  \centering
  \includegraphics[width=85mm]{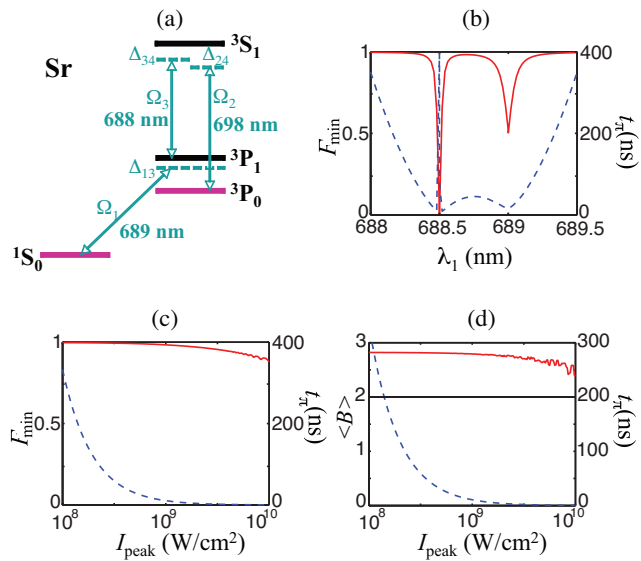}
  \caption{(Colour online) (a) Energy levels of Sr and three-photon transition for manipulation of the qubit encoded in $^1S_0$ and $^3P_0$. (b) Minimum fidelity $F_\mathrm{min}$ of single qubit operation in Sr (solid red line) and time scale for $\pi$ pulse (dashed blue line) as a function of $\lambda_1\,(=\lambda_3)$ using a peak laser pulse irradiance of $10^9\;\mathrm{W/cm}^2$. $\lambda_2$ is determined by the on-resonance condition for the three-photon transition. (c) Minimum fidelity $F_\mathrm{min}$ of single qubit operation (solid red line) and time scale for $\pi$ pulse (dashed blue line) as a function of laser irradiance $I_\text{peak}$. Detuning is fixed to $\lambda_1=688.7$~nm. (d) Resulting expectation value of the Bell operator $ \langle B \rangle$ and threshold for a local hidden variable model (solid black line).}
  \label{fig2}
\end{figure}

Recent attempts to cool and trap neutral Yb and Sr have been very successful, and we therefore consider them primarily. Optical clock states in Yb and Sr have extremely low decoherence rates, due to the fact that electric dipole one- and two-photon transitions between $^1S_0$ and $^3P_0$ states are dipole and parity-forbidden, respectively [see Figs.\ref{fig2}(a) and \ref{fig3}(a)for energy levels and transition wavelengths]. While affording long lifetimes, the selection rules also present a significant challenge to fast coherent manipulation and measurement of qubits. To overcome this challenge, we employ a coherent, three-photon transition to perform single qubit operations, utilizing the excited $^3S_1$ and $^3P_1$ states~\cite{Hong:2005}. The three transitions  $^1S_0 \rightarrow\, ^3P_1$, $^3P_1 \rightarrow\, ^3S_1$, and $^3S_1 \rightarrow\, ^3P_0$ are electric-dipole allowed (see~\cite{Porsev:1999,Werij:1992} for transition matrix elements). Because three beams can always be arranged in a plane such that the transferred recoil cancels, this three-photon transition has the benefit of being recoil-free~\cite{Hong:2005}. For Sr, the need for three lasers may be reduced to two, as explained below.

We model this three-photon transition by a master equation using the Liouvillian matrix given in~\cite{Hong:2005}. Its fidelity is limited by the short-lived intermediate $^3S_1$ state, which decays primarily to the $^3P_1$ state. The fast coherent rotation of qubits is followed by the fast readout of the $^3P_0$ state via REMPI on a nanosecond or even picosecond time scale. Re-using the $^3S_1$ excited state, photoionization can then be accomplished in a two-step process. An on-resonant $^3P_0$ to $^3S_1$ transition is followed by a final ionization step at  $\lambda<563$ nm for Yb and $\lambda<592$ nm for Sr. The main errors in this read-out scheme are due to population in the $^3P_1$ to $^3S_1$ states. During readout, any population in $^3P_0$ and $^3S_1$ will be counted as logical $\left| 1 \right\rangle$ (ionized). Population in $^1S_0$ and $^3P_1$ will be counted as logical $\left| 0 \right\rangle$ (not ionized).

The case of Sr is particularly interesting: the transitions $^1S_0 \rightarrow ^3P_1$ and $^3P_1 \rightarrow ^3S_1$ are close in energy difference (689 and 688 nm, respectively) so that a resonant two-photon transition $^1S_0 \rightarrow ^3S_1$ utilizing a single laser is possible. This reduces the laser requirement from three to two. Figure~\ref{fig2}(b) shows fidelities for qubit rotation for wavelengths in the range 688 to 689.5 nm. The time for a $\pi$-rotation is minimized by tuning to 688.7 nm. Figure~\ref{fig2}(c) shows the fidelity and time scales for a $\pi$-rotation as a function of laser powers. For fairly realistic mode-locked laser powers, $10^9\,\mathrm{W/cm}^2$ (roughly 1 kW pulse peak power focused onto $100\mathrm{\mu m}^2$), rotations within a few nanoseconds are possible with better than $90\%$ fidelity. Higher fidelities of $99.99\%$ can be reached for the same detuning by using lower laser powers of $10^6\,\mathrm{W/cm}^2$.


\section{Tests of local realism}
We show the efficacy of our fast measurement scheme by applying it to a test of local realism. This is expressed in the usual Clauser-Horne-Shimony-Holt (CHSH) form~\cite{Clauser:1969},
\begin{equation}
\label{bell}
\left\langle B \right\rangle =  \left\langle QS \right\rangle + \left\langle RS \right\rangle + \left\langle RT \right\rangle - \left\langle QT \right\rangle \leq 2,
\end{equation}
for local realistic theories, whereas Tsirelson's quantum upper bound is~$2\sqrt{2}$. For a $|\Psi^+\rangle$ entangled state, the quantum bound is saturated for $Q=Z$, $R=X$, $S=(X-Z)/\sqrt{2}$, and $T=(X+Z)/\sqrt{2}$, with $X$, $Y$, $Z$ the Pauli operators. These measurements are obtained via basis rotations  $\mathcal{R}(\theta) = \exp(+i\theta Y/2)$ applied to the state, followed by measurements in the $z$-basis. This corresponds to measurements of the form $Q=U_Q^\dagger Z U_Q$ with $U_Q=\openone$, $U_R=\mathcal{R}(\pi/2)$, $U_S=\mathcal{R}(3\pi/4)$, and $U_T=\mathcal{R}(\pi/4)$.
\begin{figure}[h]
  \centering
  \includegraphics[width=85mm]{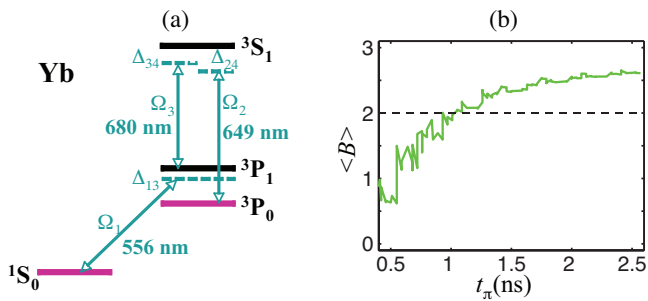}
  \caption{(Colour online) (a) Energy levels of Yb and three-photon transition for manipulation of the qubit encoded in $^1S_0$ and $^3P_0$. (b) Expectation value of the Bell operator for imperfect single-qubit rotations in Yb as a function of time scale of the measurement for $I_\mathrm{peak}=10^9\;\mathrm{W/cm}^2$.}
  \label{fig3}
\end{figure}

Inequality~(\ref{bell}) is tested by first preparing an entangled Bell state via a controlled phase gate as discussed above, then separating the atoms by a few meters. In a far-off-resonance, magic-wavelength trap, qubit coherence times are on the order of 10~s or longer. For accelerations of 200~$\text{mm}/\text{s}^2$ or faster~\cite{Gustavson:2001}, separations of a few meters should be feasible.  At this distance, synchronous measurements on a nanosecond time scale are required to ensure space-like separation.  Within this time window, the measurement basis is chosen randomly, qubits are rotated to reflect the choice of measurement basis, and qubit states are measured in the computational basis using REMPI. Fast random basis selection can be accomplished by using a light emitting diode (LED) as in~\cite{Weihs:1998}. The time necessary for this random basis selection can be minimized [e.g., by using shorter signal paths and custom-built electro-optic modulators (EOMs)] to ensure basis selection times of less than 10~ns. Rotation of the measurement basis is achieved via a coherent coupling of the qubit states via three-photon Raman transitions. The presence of the ion (i.e., the freed electron) will be detected via a single channel electron multiplier with above $99\%$ efficiency~\cite{Hurst:1979} .

As in a typical single channel experiment~\cite{Rowe:2001}, the measurement outcome can be only ``ion''$\equiv |1\rangle$ or ``no ion''$\equiv |0\rangle$. No data are discarded, and no assumptions are made about ``fair sampling''~\cite{Clauser:1969} or ``enhancement''~\cite{Clauser:1974}. Loss of an atom will result in a ``no ion''$\equiv |0\rangle$ count, which reduces the degree of Bell inequality violation but does not open any loopholes. High transport and detector efficiencies are necessary to ensure that a violation occurs. A calculation of the CHSH-type Bell inequality violation~\cite{Clauser:1969}, including errors in rotation and ionization readout, is shown in Fig.~\ref{fig2}(d) for Sr and Fig.~\ref{fig3}(b) for Yb. To achieve an average value of the Bell-operator larger than 2, as required for a violation, measurements on a time scale of a few nanoseconds (including signal processing times) should be possible with either atomic species.


\section{Conclusions} We propose schemes for fast recoil-free manipulation and measurement of qubits in Sr or Yb, and discuss an entangling operation for identical bosons in optical tweezers based on the exchange interaction first discussed for fermions in~\cite{Hayes:2007}. We furthermore show that it is possible to simultaneously close both space-like separation and detection loopholes for group-II-like atomic qubits separated on only a laboratory scale. This lays the groundwork for future exploration of measurement-based computation. Finally, our work identifies major challenges and provides concrete guidelines for experiments utilizing bosonic Yb or Sr for quantum information processing applications. 


\begin{acknowledgments}
We thank P. Julienne and A. Derevianko for helpful discussions on Yb and Sr, W. Ketterle for discussions on long-distance transport of atoms, D. Hayes and I. Deutsch for discussions on entangling atoms via exchange interactions, and K. Resch and G. Weihs for discussions on Bell inequalities. This work was supported by NSERC, AIF, CIFAR, iCORE, MITACS, NSF, and The Welch Foundation.
\end{acknowledgments}


\end{document}